# ZnO Nanocrystal Networks Near the Insulator-Metal Transition: Tuning Contact Radius and Electron Density with Intense Pulsed Light

*Benjamin L. Greenberg,[a*] Zachary L. Robinson,[b] K. V. Reich,[b,d] Claudia Gorynski,[e] Bryan N. Voigt,[c] Lorraine F. Francis,[c] B. I. Shklovskii,[b] Eray S. Aydil,[c] and Uwe R. Kortshagen[a*]*

[a]Department of Mechanical Engineering, [b]Department of Physics, and [c]Department of Chemical Engineering and Materials Science, University of Minnesota, Minneapolis, Minnesota 55455, United States

[d]Ioffe Institute, St Petersburg, 194021, Russia

[e]Department of Mechanical and Process Engineering, University of Duisburg-Essen, Forsthausweg 2, 47057 Duisburg, Germany

ABSTRACT. Networks of ligand-free semiconductor nanocrystals (NCs) offer a valuable combination of high carrier mobility and optoelectronic properties tunable via quantum confinement. In principle, maximizing carrier mobility entails crossing the insulator-metal transition (IMT), where carriers become delocalized. A recent theoretical study predicted that this transition occurs at $n\rho^3 \approx 0.3$, where $n$ is the carrier density and $\rho$ is the interparticle contact radius. In this work, we satisfy this criterion in networks of plasma-synthesized ZnO NCs by using intense pulsed light (IPL) annealing to tune $n$ and $\rho$ independently. IPL applied to as-deposited NCs increases $\rho$ by inducing sintering, and IPL applied after the NCs are coated with Al$_2$O$_3$ by atomic layer deposition increases $n$ by removing electron-trapping surface hydroxyls. This procedure does not substantially alter NC size or composition and is potentially applicable to a wide variety of nanomaterials. As we increase $n\rho^3$ to at least twice the predicted critical value, we observe conductivity scaling consistent with arrival at the critical region of a continuous quantum phase transition. This allows us to determine the critical behavior of the dielectric constant and electron localization length at the IMT. However, our samples remain on the insulating side of the critical region, which suggests that the critical value of $n\rho^3$ may in fact be significantly higher than 0.3.

Understanding and controlling charge transport across semiconductor nanocrystal (NC) networks has enabled NC-based devices such as solar cells,[1,2] thermoelectric generators,[3,4] and field-effect transistors.[5,6] The NCs are typically covered with and separated by ligands, whose composition and length can be varied to tune carrier mobility across several orders of magnitude.[7–9] Alternatively, NCs have been brought into direct contact by removing these ligands[10–13] or by using gas-phase synthesis and deposition,[14–17] and mobility has been tuned by varying the interparticle contact area.[10,12,14,18] Elimination of ligands promises particularly high mobility, and envisioned applications of this approach prompt a fundamental question: what are the necessary conditions for metallic transport?

An insulator-metal transition (IMT) criterion for direct-contact NC networks (DCNNs),

$$(n\rho^3)_c \approx 0.3g, \qquad\qquad (1)$$

was proposed,[19] where (for an n-type semiconductor) $n$ is the free electron density, $\rho$ is the contact radius between adjacent NCs, and $g$ is the degeneracy of the conduction band minima (for ZnO, $g = 1$). This derives from the requirement at the IMT that the interparticle contact conductance, $G_i$, is equal to the quantum conductance, $e^2/\pi\hbar$,[20,21] which is satisfied when the free electron Fermi wavelength, $k_F^{-1} = (g/3\pi^2 n)^{1/3}$, is comparable to $\rho$. As shown in Figure 1a, when $\rho \approx 1$ nm, Equation 1 predicts that the critical value of $n$ is on the order of $10^{20}$ cm$^{-3}$, which for 10-nm-diameter NCs corresponds to ~50 free electrons per NC. Hence seven electron shells are occupied and $\epsilon_F = \hbar^2 k_F^2/2m^*$ is a good approximation of the highest occupied shell energy.[19]

Experimental efforts to achieve metallic transport in DCNNs and to verify Equation 1 have yielded promising yet inconclusive results. Chen *et al.*[19] measured the conductivity of thin films of plasma-synthesized P-doped Si NCs. Increasing $n$ so that so that $n\rho^3$ reached approximately half the predicted critical value resulted in electron localization length, $\xi$, greater than the NC diameter, $d$, as expected near the IMT. More recently, higher $n\rho^3$ values were achieved in networks of plasma-synthesized ZnO NCs.[14] While $n$ was held constant, $\rho$ was increased via atomic layer deposition (ALD) of additional ZnO onto the NC surfaces, yielding $n\rho^3$ greater than the predicted critical value. Some evidence of an IMT was given: the most conductive sample exhibited $\xi > d$, $G_i > e^2/\pi\hbar$, and $n$ independent of temperature down to 100 K. The temperature coefficient of conductivity, however, remained positive, and $n$ was not measured below 100 K.

These results reveal a need for further study of not only the applicability of Equation 1 but also of the critical behavior of electrical properties near the IMT in DCNNs. Proximity to the IMT has been difficult to evaluate, because knowledge of the signals of the transition in DCNNs has been relatively limited. For bulk semiconductors, it is well known that the IMT is often a continuous quantum phase transition with a critical region in which conductivity attains a power-law dependence on temperature, and $\xi$ and the dielectric constant, $\varepsilon$, diverge as powers of the tuning parameter that drives the transition.[22–24] To our knowledge, for DCNNs, such a transition has not been demonstrated and such dependencies have not been established.

In this work we test Equation 1 and examine in detail the near-IMT transport behavior of DCNNs by taking a new approach to manipulating $n\rho^3$. We develop a technique for independently tuning both $\rho$ and $n$ in a ZnO DCNN while maintaining constant $d$ and network composition, so that we can isolate the dependencies of electrical properties on $n\rho^3$ throughout the critical range. This approach is distinct from impurity doping, which introduces impurity phases that may impede transport,[19,25] and from the ZnO ALD method which significantly increases $d$ and ZnO volume fraction in addition to $\rho$. The ability to modulate both $\rho$ and $n$ without changing $d$ is useful not only for this study but also for pursuing the goal of "confined but connected" NCs which promise wide tunability of optical as well as electrical properties.[11,13] Our technique is based on xenon-flashlamp intense pulsed light annealing (IPL), which has the additional advantages of roll-to-roll processing compatibility and applicability to a wide variety of materials,[26–30] including ZnO.[31,32] Applying IPL to a ZnO DCNN with $d = 10$ nm, we increase $\rho$ from 1.1 to 1.5 nm ($\rho^3$ from 1.5 to 3.4 nm$^3$) and then increase $n$ from 1.1 to $7.6 \times 10^{20}$ cm$^{-3}$ in order to tune $n\rho^3$ from 0.17 to 2.6, according to estimates of $\rho$ and $n$ based on ellipsometry and Fourier Transform infrared (FTIR) spectroscopy, respectively. For large $n\rho^3$, the dependence of conductivity on temperature transitions from a stretched exponential form to a power law, which suggests that the IMT critical point, $(n\rho^3)_c$, lies near the top of our measured range. Indeed, assuming $n\rho^3_c \approx 2.6$, we find that for $n\rho^3$ near $(n\rho^3)_c$, both $\zeta$ and $\varepsilon$ diverge as $[(n\rho^3)_c - n\rho^3]^{-1}$, and conductivity exhibits scaling behavior consistent with arrival at the critical region of a continuous quantum phase transition. These results provide clear measures of proximity to the IMT, and they suggest that for nonideal DCNNs such as ours and those studied by Chen *et al.*[19] and Lanigan and Thimsen[14], the criteria for crossing to the metallic side may in fact be more stringent than Equation 1.

We produce ZnO DCNNs by nonthermal plasma synthesis integrated with inertial impact deposition, as described previously.[15] Briefly, diethylzinc and oxygen gas are injected into a low-pressure (5 Torr) radio-frequency argon plasma to form ZnO NCs with $d = 10$ nm. At the outlet of the plasma chamber, the NCs pass through a slit orifice and enter a lower-pressure (10 mTorr) deposition chamber, forming a two-dimensional supersonic particle beam. We translate substrates perpendicularly to the beam to create $1 \times 1$ cm ZnO DCNN films with thickness, $t$, of $300 \pm 20$ nm and ZnO volume fraction, $\phi$, of $33 \pm 2$ %. The NCs in these films are ligand-free and in direct contact, and their surfaces are terminated in hydroxyl and carboxylate groups. The hydroxyl groups are electron traps, so we remove them by infilling the films' pores with $Al_2O_3$ by ALD,[15] which yields air-stable $n$ on the order of $10^{20}$ cm$^{-3}$ (see Note 2 in the Supporting Information for further discussion). The NCs are not intentionally doped, and the high $n$ is likely due to oxygen vacancies,[33,34] although we cannot rule out other donors such as hydrogen[35] and interstitial zinc.[36]

To develop a strategy for independently tuning $\rho$ and $n$ while maintaining constant $d$ and network composition, we consider the following. First, we can target $\rho$ by encouraging the NC network to enter – but not surpass – the initial stage of sintering. During this initial stage, grains minimize surface energy by forming necks, but otherwise grains retain their shape, and substantial grain growth does not occur[37–39] (see Figure S1 in the Supporting Information). This can be accomplished by adding an annealing step before the ALD infilling which thermally activates solid state diffusion of NC atoms toward the interparticle contacts. Then, to modulate $n$, we can manipulate the composition of the film after ALD. Specifically, we can exploit the fact that after the NC network is infilled with $Al_2O_3$, the film still contains some electron-trapping hydroxyl groups due to incomplete removal and/or introduction of new hydroxyls during the ALD process.[40] Reducing the concentration of these residual traps should increase $n$.

We realize this strategy by introducing IPL both before and after the ALD step. In both cases, we apply IPL under ambient conditions at a surface power density of 12 kW/cm$^2$ with a 660 ms period and a 1 ms flash duration. The ZnO NCs strongly absorb the UV portion of the xenon flashlamp spectrum. In general, absorption of a UV photon with energy greater than the band gap of ZnO (3.3–3.4 eV) alters the surface chemistry of ZnO by a two-step process: the photon generates an electron-hole pair, and then the hole oxidizes an electronegative and/or negatively charged surface species, resulting in desorption.[31,41–43] The remaining photogenerated electron can persist as a free charge carrier or it can be trapped upon readsorption of the desorbed species or adsorption of another oxidant. When we apply IPL before ALD, these processes induce sintering: electron and hole relaxation to the band edges and non-radiative electron-hole recombination supply thermal energy, and photochemical removal of stabilizing surface species facilitates formation of new Zn-O bonds. We will show shortly that removal of surface carboxylates is particularly important. However, while $\rho$ increases, $n$ remains constant due to adsorption of ambient gases. In contrast, when we apply IPL after ALD, the $Al_2O_3$ infill suppresses sintering so that $\rho$ remains constant. Meanwhile hydroxyl groups are removed, and the $Al_2O_3$ infill protects the photogenerated electrons from oxidizing gases so that $n$ increases.

Figures 1b, 1c, and 1d show our treatment sequence and summarize its effects on $\rho$, $n$, and NC surface chemistry. We combine IPL before ALD ("sintering IPL") and IPL after ALD ("doping IPL") to produce two data series: (1) a sintering series (varied $\rho$, constant $n$) in which the number of sintering IPL flashes is varied from 0 to 1000 and (2) a doping series (varied $n$, constant $\rho$) in which all samples are pretreated with 1000 flashes of sintering IPL, and then the number of doping IPL flashes is varied from 1 to 1000. Sample names are of the form "$x + y$", where $x$ is the number of sintering IPL flashes, and $y$ is the number of doping IPL flashes. Neither sintering IPL nor doping IPL significantly alters $d$ or film microstructure, as indicated by X-ray diffraction (XRD) patterns and scanning electron micrographs (SEM) (Figure S2 in the Supporting Information).

Figure 2a shows the increase in $\rho^3$ with the number of sintering IPL flashes. To detect changes in $\rho$ we measure film thickness, $t$, and ZnO volume fraction, $\phi$, by spectroscopic ellipsometry, using the Bruggeman effective medium approximation. We perform the measurements before ALD, so that there is no $Al_2O_3$ coating and no significant contribution of free electrons to the dielectric function ($n \approx 0$). This approach was applied to similar ZnO NC networks previously and was shown to yield $t$ and $\phi$ values in good agreement with scanning electron microscopy and Rutherford backscattering data.[15] Representative spectra are shown in Figure 2b, and a representative fit and complete plots of $t$ and $\phi$ are shown in Figure S3 in the Supporting Information. During sintering, the ZnO NC film slightly shrinks and densifies: 1000 flashes of sintering IPL decreases $t$ by ~6% and increases $\phi$ by ~6%. These modest changes are consistent with remaining in the initial stage of sintering rather than proceeding to the intermediate stage (characterized by more substantial densification and changes in grain morphology).[39] We convert changes in $\phi$ to changes in $\rho$ using a simple geometrical model described in Note 1 in the Supporting Information. In this model, we treat $\rho$ as homogeneous throughout the network, and we must make an assumption about the starting value of $\rho$ in an unsintered network, $\rho_0$. We assume $\rho_0 = 0$. In reality, an unsintered network likely contains a mix of interfacial and point contacts, so that our model may lead to conservative estimates of $\rho$. However, we expect sintering to improve the accuracy of the homogeneity assumption, because surface area minimization should entail reduction of $\rho$ dispersion.

Since all conductivity measurements are performed after the ALD infilling step, it is important to estimate $\rho$ after ALD. The ALD temperature is 180 °C, so some additional sintering may occur, increasing the final value of $\rho$. Figure 2b shows $\rho^3$ both before and after ALD. To obtain $\rho$ before ALD, we conduct ellipsometry measurements immediately after sintering IPL, and we find that $\rho^3$ increases from 0 to 2.6 ± 0.8 $nm^3$ as we increase the number of sintering IPL flashes from 0 to 1000. Then, to obtain $\rho$ after ALD, we heat the samples under $N_2$ at 180 °C for 33 minutes in order to mimic the temperature history of ALD-treated samples without actually depositing $Al_2O_3$, since an $Al_2O_3$ infill would compromise ellipsometry data fitting. This heating does in fact induce additional sintering (without grain growth) and reveals that $\rho$ after ALD is approximately 0.8 to 1.5 nm larger than $\rho$ before ALD: after heating, $\rho^3$ ranges from $1.5 \pm 0.3$ nm$^3$ (0 sintering IPL flashes) to $3.4 \pm 0.8$ nm$^3$ (1000 sintering IPL flashes). We use these post-ALD $\rho^3$ estimates to determine $n\rho^3$.

Post-ALD $\rho$ appears to depend on carboxylate removal during sintering IPL, which is evident in the Fourier transform infrared (FTIR) spectra (Figure 3a). The absorption feature at 1600 cm$^{-1}$, which we attribute to surface carboxylates,[44–47] vanishes as we increase the number of sintering IPL flashes. This removal of carboxylates is likely due to oxidation by UV-photogenerated holes, as illustrated in Figure 3a. The importance of carboxylate removal was revealed by a control experiment: with a low-intensity Xe lamp, we illuminated a sample with a total photon dose equivalent to 100 flashes of sintering IPL but at a much lower fluence, so that carboxylate removal was the only observed effect and no sintering occurred during illumination (see Figure S4 in the Supporting Information). We found that pre-ALD $\rho^3$ was unchanged, but that post-ALD $\rho^3$ was higher than that of a non-illuminated sample and equal to (within 3% of) that of a sample treated with 100 flashes of sintering IPL. (Also, the same was true of electrical conductivity.) This surprising result suggests that carboxylate removal is in fact the most important effect of sintering IPL, because it determines the degree of additional sintering during ALD and thus the final value of $\rho^3$. A more detailed explanation of the role of carboxylates in sintering is proposed in Figure S5 in the Supporting Information.

Figure 3b shows the FTIR spectra of the doping series (acquired after ALD), which reveal hydroxyl removal and the resultant increase in $n$. The absorption feature at ~3500 cm$^{-1}$ that vanishes with doping IPL is due to hydroxyl groups.[15,48] As mentioned earlier, hydroxyls may be present on the ZnO NC surfaces due to incomplete removal by ALD, or they may be introduced during ALD. A hydroxyl originating from ALD and residing in the $Al_2O_3$ is likely to act as a trap if its distance from an NC surface is less than the decay length of the electron wavefunction in $Al_2O_3$. As shown in the schematic in Figure 3b, we attribute removal of hydroxyls on or near the ZnO NC surfaces to oxidation by photogenerated holes (see figure S4 and Note 2 in the Supporting Information for additional data and discussion). The broad and intense absorption features centered near ~2000 $cm^{-1}$ in Figure 3b are due to localized surface plasmon resonance (LSPR). That they blueshift and intensify with doping IPL confirms that hydroxyl removal is accompanied by rising $n$.

We quantify $n$ by fitting the LSPR absorption features. We favor this approach over Hall effect measurements (see Figure S6 in the Supporting Information), which produce results that are difficult to interpret in a consistent manner for a diverse set of samples exhibiting hopping conduction[49,50] and approaching the IMT. Beginning with the last sample of the sintering series, "1000+0" (1000 sintering IPL flashes, 0 doping IPL flashes), we estimate the film's dielectric function using the Maxwell Garnett effective medium approximation (MG EMA). The MG EMA, which does not account for interparticle conduction, has previously been shown to produce good fits of high-$n$ films of $In_2O_3$:Sn NCs exhibiting hopping transport, and has been found to outperform the Bruggeman EMA[51] (whereas the Bruggeman EMA was shown to produce accurate fits of ellipsometry data from films of ZnO NCs with $n \approx 0$,[15] in which the dielectric function is determined by bound electrons). In the MG EMA, the overall effective dielectric function of the film, $\varepsilon$, is given by

$$\frac{\varepsilon - \varepsilon_m}{\varepsilon + 2\varepsilon_m} = \phi \frac{\varepsilon_{NC}(\omega) - \varepsilon_m}{\varepsilon_{NC}(\omega) + 2\varepsilon_m} \qquad (2)$$

where $\phi$ is the NC volume fraction, $\varepsilon_m$ is the dielectric constant of the medium in which the NCs are embedded ($Al_2O_3$, $\varepsilon_m = 2.6$ at high frequency[52]), and $\varepsilon_{NC}(\omega)$ is the frequency-dependent dielectric function of the NCs given by the Drude equation, $\varepsilon_{NC}(\omega) = \varepsilon_\infty - \omega_p^2/(\omega^2 + i\omega\Gamma)$, where $\varepsilon_\infty$ is the high-frequency dielectric constant (3.7 for ZnO[52]), $\Gamma$ is the carrier relaxation frequency, and $\omega_p$ is the plasma frequency. We approximate $\Gamma$ as a constant $\Gamma = e/\mu_l m^*$, and $\omega_p$ is given by $\omega_p^2 = ne/\varepsilon_0 m^*$, where $e$ is the elementary charge, $\mu_l$ is the local electron mobility within an NC (which we use as a fitting parameter), $m^*$ is the electron effective mass ($0.3 m_e$ for ZnO[52]), and $\varepsilon_0$ is the permittivity of free space. A fit of the "1000+0" spectrum to the MG EMA absorption obtained from the imaginary part of $\varepsilon$, shown in Figure 3b, gives $n = 1.1 \times 10^{20}$ cm$^{-3}$ (and $\mu_l = 21$ cm$^2$V$^{-1}$s$^{-1}$, in agreement with the value previously reported for plasma-synthesized ZnO NCs[15]). For samples approaching the IMT, however, it is less clear how to select or formulate an appropriate EMA and whether the same EMA should be used as transport behavior evolves. Therefore, for the doping series, rather than attempting to model interparticle interactions, we treat the network as a single absorber and determine $n$ from the area under the LSPR absorption feature, which is proportional to $n$.[53] Using $n = 1.1 \times 10^{20}$ cm$^{-3}$ as a baseline, we find from the increased area that $n$ rises to a maximum value of $7.6 \times 10^{20}$ cm$^{-3}$ after 1000 flashes of doping IPL. This is similar to the maximum $n$ attained in bulk impurity-doped ZnO[54,55] and in ZnO NCs photodoped using borohydride hole quenchers[56] ($\sim 10^{21}$ cm$^{-3}$ and $\sim 6 \times 10^{20}$ cm$^{-3}$, respectively) and translates to a maximum $n\rho^3$ of 2.6.

Both sintering IPL and doping IPL increase the electrical conductivity of the ZnO NC network, $\sigma$, according to four-point measurements in the Van der Pauw configuration performed after ALD at 300 K (Figure 4a). Without any IPL, $\sigma = 3.8 \times 10^{-2}$ Ω$^{-1}$cm$^{-1}$. 1000 flashes of sintering IPL increases $\sigma$ by two orders of magnitude, so that the "1000+0" sample has $\sigma = 3.7\ \Omega^{-1}\text{cm}^{-1}$. Then 1000 subsequent flashes of doping IPL increases $\sigma$ by an additional factor of five, so that the "1000+1000" sample has $\sigma = 1.8 \times 10^1\ \Omega^{-1}\text{cm}^{-1}$. This is merely a factor of ten less than $\sigma$ of typical dense microcrystalline ZnO thin films without intentional doping,[54] and is the highest $\sigma$ we achieved in our ZnO NC networks by IPL, i.e., additional doping and sintering IPL flashes yielded insignificant benefit. That a combination of sintering and doping IPL is necessary for attaining this value was confirmed by control experiments in which samples were treated with 2000 flashes of either sintering or doping IPL alone (see Table S1 in the Supporting Information). Figure 4b demonstrates that $n$ and network electron mobility, $\mu$, are independently tunable by doping and sintering IPL, respectively, where $n$ is determined from the LSPR fits and $\mu = \sigma/en$. Sintering IPL does not significantly change the LSPR (see Figure S7 in the Supporting Information), which indicates that $n$ remains constant, and that the increase in $\sigma$ is due entirely to enhancement of $\mu$. In contrast, doping IPL increases $n$ while $\mu$ remains nearly constant, which is consistent with our expectation that doping IPL does not increase $\rho$, the primary determinant of $\mu$. Although the $\text{Al}_2\text{O}_3$ infill precludes measuring $\rho$ of the doping series samples by ellipsometry, we can now justifiably assume that $\rho$ is constant when calculating the corresponding values of $n\rho^3$.

We can use $\sigma$ of the ZnO NC network to estimate the interparticle contact conductance, $G_i$, (right axis of Figure 4a).[14] We treat the NC network as a random resistor network and neglect the small contribution of intraparticle resistance, i.e., we assume that the resistance of the network is due to interparticle contact resistance only, a reasonable approximation for a network with $\mu$ much lower than that of bulk ZnO. In this case, according to the three-dimensional nodes and links model,[57]

$$G_i = \frac{\sigma d}{(\phi - \phi_0)^{1.9}}, \tag{3}$$

where $\phi_0$ is the percolation threshold (the minimum ZnO volume fraction at which percolation occurs), which was found to be 5% for our ZnO NC networks.[14] From Equation 3, we estimate $G_i = 1.9 \times 10^{-4}\,\Omega^{-1}$ for the "1000+1000" sample, which is greater than the quantum conductance, $e^2/\pi\hbar = 7.7 \times 10^{-5}\,\Omega^{-1}$, suggesting proximity to the IMT.

Having established that $n\rho^3$ surpasses the predicted critical value of ~0.3, and that $G_i$ exceeds $e^2/\pi\hbar$, we proceed to examine the temperature dependence of $\sigma$ near the IMT. Figure 5a demonstrates that for less conductive samples, log $\sigma$ appears linear in $T^{-1/2}$, which is consistent with the Efros-Shklovskii variable range hopping (ESVRH) law,

$$\sigma = \sigma_0 \exp\left[-(T_{ES}/T)^{1/2}\right], \tag{4}$$

with $T_{ES}$ given by

$$T_{ES} = \frac{Ce^2}{\varepsilon k_B \xi}, \tag{5}$$

where $e$ is the elementary charge, $k_B$ is the Boltzmann constant, $\varepsilon$ is the dielectric constant of the NC network, $\xi$ is the electron localization length, and $C \approx 9.6$[58] is a numerical coefficient. ESVRH is confirmed by a Zabrodskii plot of the logarithmic derivative $W = d[\ln(\sigma)]/d[\ln(T)]$ against $T$ on a logarithmic scale[49,59] (Figure S8 in the Supporting Information), in which the dynamic range of conductivity is ~$10^4$. This analysis reveals, however, that the "1000+1000" sample does not exhibit ESVRH, and that signs of deviation from ESVRH emerge in the "1000+100" sample. Indeed, as $n\rho^3$ increases, $T_{ES}$ (extracted from the slope of log $\sigma$ vs. $T^{-1/2}$) declines to values at which hopping is expected to give way to metallic transport (Figure 5b). In an insulating NC network, we expect $\varepsilon$ to be less than that of bulk ZnO (~9) and $\xi$ to be less than $d$ (10 nm), which translates to $T_{ES} \gtrsim 2000$ K (a theoretical estimate of $\xi$ from $n\rho^3$ is given in Note 3 in the Supporting Information). This inequality is satisfied for $n\rho^3 \approx 0.2$, whereas for large $n\rho^3$, $T_{ES}$ falls to 40 K before becoming indeterminate.

The temperature dependence of $\sigma$ in our most conductive sample is consistent with a continuous quantum IMT. In such transitions in bulk semiconductors, stretched exponential behavior is supplanted by a power law at the IMT at low temperature.[23,24] This shift can be explained by the condition that the critical conductivity at the IMT is $\sigma_c = e^2/\pi\hbar L_\varphi$, where $L_\varphi$ is the electron phase relaxation length. The temperature dependence of $L_\varphi$ is determined by the electron-electron interaction energy,[60] which at the IMT is inversely proportional to a power, $z$, of distance. It has been shown that $z = 2$,[61,62] which leads to $L_\varphi \sim T^{-1/2}$ and thus $\sigma_c \sim T^{1/2}$. Figure 5c shows that our most conductive sample, "1000+1000" ($n\rho^3 = 2.6$), exhibits a temperature dependence of $\sigma$ close to this power law. This is supported by the Zabrodskii plot in Figure S8, which shows that $W$ at low temperature is nearly constant and approaches 1/2, suggesting that the distance from the critical point is nonzero but small enough that we may hypothesize that $(n\rho^3)_c \approx 2.6$.

The scaling behavior of $\sigma$ (Figure 5d) further supports arrival at the critical region of a continuous quantum IMT and allows us to determine the critical indices of the transition. Near an IMT of this type, if $n\rho^3$ is the tuning parameter that drives the transition,[63,64] then $\xi$ should diverge as $[\Delta(n\rho^3)]^{-\nu}$, where $\Delta(n\rho^3) = (n\rho^3)_c - n\rho^3$, and $\nu$ can be determined empirically. Also, $\sigma$ should scale with $\xi/L_\varphi$ so that

$$\sigma = \sigma_c f\left[\Delta(n\rho^3)/T^{1/vz}\right], \tag{6}$$

where $f(x)$ is the scaling function. Typically the function $f(x)$ has two branches, so that conductivity data collapse into two curves on opposite sides of the transition.[23] With our data we can check for this behavior on the insulating side. Figure 5d shows that, when we take $(n\rho^3)_c = 2.6$ (i.e., "1000+1000" is the critical sample), the conductivities of the other four most conductive samples ("1000+0",…, "1000+100") follow Equation 6 with $v = 1$. Additionally, the inset of Figure 5d shows that for these samples, $T_{ES} \sim [\Delta(n\rho^3)]^2$. Hence we conclude that near the IMT both $\xi$ and $\varepsilon$ diverge as $[\Delta(n\rho^3)]^{-1}$.

The samples adhering to Equation 6 are precisely those treated with 1000 flashes of sintering IPL. Apparent deviation from Equation 6 in samples not fully sintered could be explained simply by greater distance from the critical point, but it is worth noting that in these samples, scaling may be difficult to detect due to error in our estimates of $\Delta(n\rho^3)$. As mentioned earlier, contact radius dispersion in these samples may reduce the accuracy of our geometrical model. If sintering does in fact improve model accuracy, then a certain number of sintering IPL flashes may be necessary for detecting scaling, regardless of distance from the critical point.

Finally, we evaluate the applicability of $(n\rho^3)_c \approx 0.3$ (Equation 1) to our DCNNs. Figures 5c and 5d are consistent with a significantly higher critical value, $(n\rho^3)_c \approx 2.6$, and Figure S8 suggests this is a conservative estimate, as discussed earlier. Furthermore, we must of course treat this estimate as a lower bound because of our current lack of measurements of samples with higher $n\rho^3$. To understand this apparent disagreement between theory and experiment, we first consider the possibility that we have overestimated $n\rho^3$ of our samples, which would most likely be due to overestimating $n$, since our estimates of $\rho$ are probably conservative, as discussed earlier. Indeed, we have indicated the difficulty of modelling interparticle interactions when extracting $n$ from the LSPR absorption features of NC networks approaching the IMT. To obtain a lower bound on $n$ with respect to these interactions, we can neglect interparticle effects altogether, i.e. we can model the NC network as a collection of isolated spheres embedded in $Al_2O_3$ using Mie theory,[65] which gives $\omega_{LSPR} \approx [ne^2/(\varepsilon_0 m^*(\varepsilon_\infty + 2\varepsilon_m))]^{1/2}/2\pi$ (where we use $\varepsilon_\infty = 3.7$ and $\varepsilon_m = 2.6$ as before). For the critical sample, this approach yields $n \approx 2 \times 10^{20}$ cm$^{-3}$ instead of $7.6 \times 10^{20}$ cm$^{-3}$, which translates to $(n\rho^3)_c \approx 0.7$ instead of 2.6. This lowers the disagreement with theory to at least a factor of two, which is a lower bound large enough that physical explanations are worth considering. In the model used to derive Equation 1, the DCNN is a cubic lattice of epitaxially connected NCs, whereas in reality our DCNNs have ZnO volume fraction $\phi \approx 33\%$ (close to tetrahedral packing) and the NCs are oriented randomly. The significance of these non-idealities is illuminated by an alternative derivation of Equation 1 based on the condition that $tK/\Delta = 1$ at the IMT,[64] where $t$ is the tunneling matrix element for neighboring NCs, $K$ is the NC network's connectivity constant, and $\Delta$ is the energy gap between consecutive electron shells. Low $\phi$ correspond to small $K$, and NC misorientation reduces $t$, both leading to higher $(n\rho^3)_c$. Another issue to consider is the possible presence of appreciable depletion regions near the NC surfaces, which may form when electrons are trapped by surface species such as hydroxyl groups. Penetration of these depletion regions into interparticle necks could reduce the effective contact area available for electron transport and thus raise the critical value of $\rho$.

In conclusion, IPL, when combined with an NC surface coating process such as ALD, is a powerful tool for tuning both $n$ and $\rho$ in a ZnO DCNN. This technique is clearly useful for fundamental charge transport studies, and it has the potential to play a key role in low-cost manufacturing of electronics based on ZnO NCs. Modulation of $\rho$ with minimal impact on $d$ should also be achievable in other UV/visible-absorbent NCs, which should enable high $\mu$ without sacrificing quantum confinement. In this work, we have used IPL to provide strong evidence of arrival at the critical region of a continuous quantum IMT and to determine the critical indices of the transition. However, even for $n\rho^3$ as high as 2.6, we have observed only the insulating side of the critical region, which suggests that satisfying Equation 1 may be insufficient for truly metallic behavior in our nonideal DCNNs. In future studies of the IMT, critical behavior should be examined at higher $n\rho^3$, and the importance of NC packing density, orientation, and surface depletion should be investigated.

METHODS. *Sample preparation*. NCs were synthesized and deposited using the nonthermal plasma reactor described previously.[15,25] The procedure was identical to that described in refs. 15 and 25 for ZnO with no intentional impurity doping. All samples were exposed to air upon removal from the reactor. NC networks were infilled with $Al_2O_3$ using a Cambridge Nanotech/Ultratech Savannah S200 ALD system and the procedure described previously.[15,25] This infilling was done before all electrical characterization and before some FTIR, XRD, and SEM measurements (specified in the figure captions). Sintering IPL and doping IPL were performed under ambient conditions using a Xenon Corp. Sinteron 2010 equipped with a 10 in. Xe U-lamp (flashlamp). The system was operated at a voltage setting of 3 kV, which translated to a surface energy density of 12 $J/cm^2$ per flash, according to the calibration provided by the manufacturer. The lamp's output waveform was approximately rectangular with a 660 ms period and 1 ms pulse width.

*Electron transport measurements.* Transport measurements were conducted on 1x1 cm samples on boro-aluminosilicate glass (Corning Eagle XG) substrates under He vapor using a Quantum Design Physical Property Measurement System (PPMS). After ALD and IPL, Al contact pads were deposited by thermal evaporation at the four corners of each square, and Au leads were attached to the Al by either wire bonding or soldering. Measurements of the temperature dependence of conductivity, $\sigma$, were performed using DC, and temperature was allowed to stabilize before each $\sigma$ measurement. At the end of each run, $\sigma$ was remeasured at room temperature to confirm that $\sigma$ did not drift significantly over time. For the doping series, current on the order of 1 to 10 μA was sourced through two adjacent contacts, which produced an Ohmic response at all temperatures, and the voltage was measured between the other two contacts using a nanovoltmeter. This method could not be applied to the sintering series, because sample resistance neared and eventually exceeded the input impedance of the nanovoltmeter. Instead, a voltage on the order of 0.1 to 1 V was applied between two adjacent contacts, which produced an Ohmic response at all temperatures, and the current between the other two contacts was measured with an electrometer. At all temperatures, contact resistance was much lower than the NC network resistance, as confirmed by comparison of two-point and four-point measurements of the doping series. Hall measurements were performed using a DC excitation of 10 μA. Further details are provided in Figure S6 of the Supporting Information.

*Other characterization.* Fourier transform infrared (FTIR) spectra were acquired under an $N_2$ atmosphere from samples on polished NaCl substrates using a Bruker Alpha IR spectrometer in transmission mode. Ellipsometry spectra were acquired from samples on boro-aluminosilicate glass (Corning Eagle XG) substrates using a J.A. Woolam M44 ellipsometer. Prior to NC deposition, the backs of the glass substrates were roughened with a grinding wheel to prevent backside reflection during measurement. Spectra were fitted in WVASE using the Bruggeman effective medium approximation, assuming two components: ZnO and void. X-ray diffraction (XRD) patterns were acquired from samples on glass and Si substrates using a Bruker D8 Discover diffractometer equipped with a Be area detector and a Co Kα source. The collected patterns were mathematically converted to Cu Kα patterns. NC diameter, $d$, was determined using the Scherrer equation with a geometrical correction factor of 4/3 to account for the NCs' spheroidal morphology.[66] Scanning electron microscopy (SEM) was performed on samples on glass substrates at an accelerating voltage of 5 keV. To minimize charging, the samples were sputter coated with 5 Å of Pt.

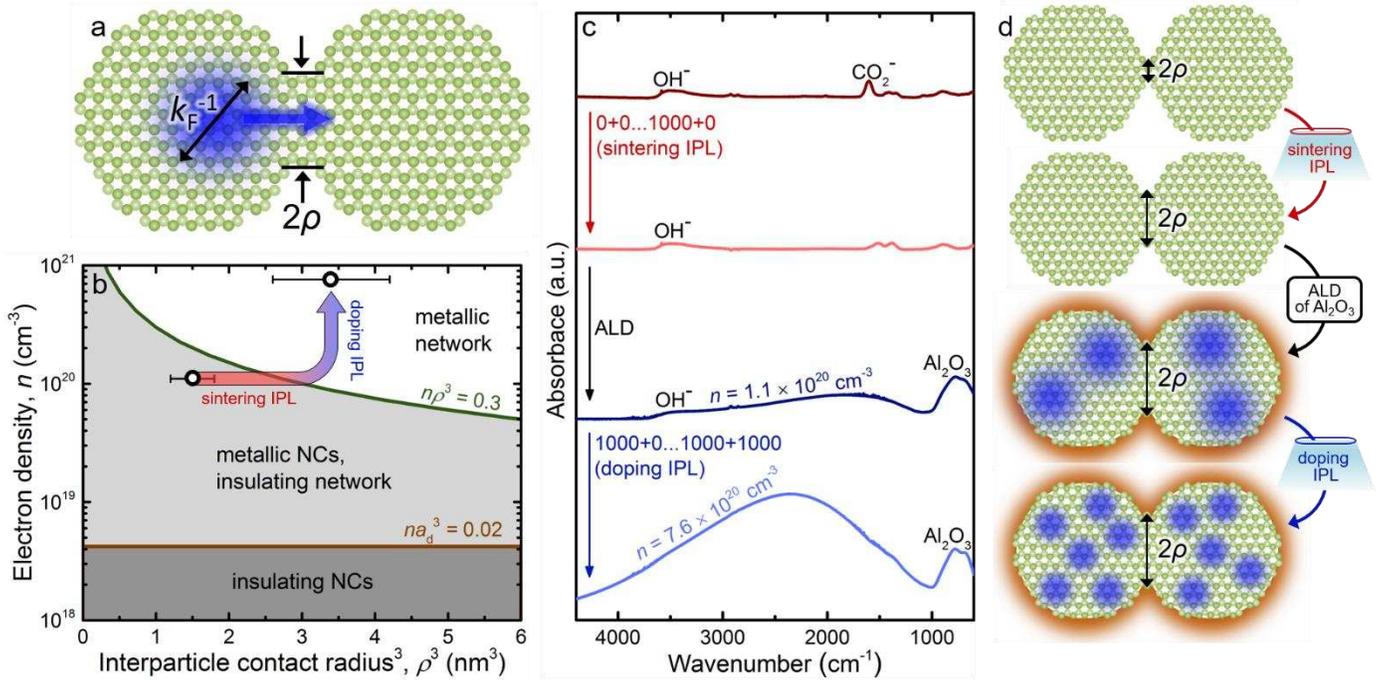

**Figure 1.** (a) Schematic of two NCs with contact radius $\rho$ and Fermi wavelength $k_F^{-1}$. Equation 1 derives from the expectation that metallic charge transport occurs for $\rho > 2k_F^{-1}$. (b) Theoretical charge transport phase diagram of a ZnO NC network. The brown line represents the Mott criterion for the bulk IMT, where $a_d \approx 1.5$ nm is the donor Bohr radius in ZnO,[67] and the green line represents the criterion for the NC network IMT (Equation 1). The data points and arrow illustrate the path taken in this work, which ends in the region where metallic transport is expected. (c) FTIR absorption spectra of ZnO NC networks at various treatment stages. Sequentially, the NCs are deposited (dark red), sintered by IPL (light red), coated with $Al_2O_3$ by ALD (dark blue), and finally photodoped by IPL (light blue). The spectra shown here correspond to maximally treated samples (1000 flashes each of sintering and doping IPL). Sample names are of the form "$x+y$", where $x$ is the number of sintering IPL flashes and $y$ is the number of doping IPL flashes. (d) Corresponding schematics of treatment stages. Sintering IPL removes surface carboxylates and increases $\rho$; then ALD removes surface hydroxyls, thereby freeing electrons (while further increasing $\rho$); then doping IPL removes remaining hydroxyls, thereby increasing $n$ and decreasing $k_F^{-1}$.

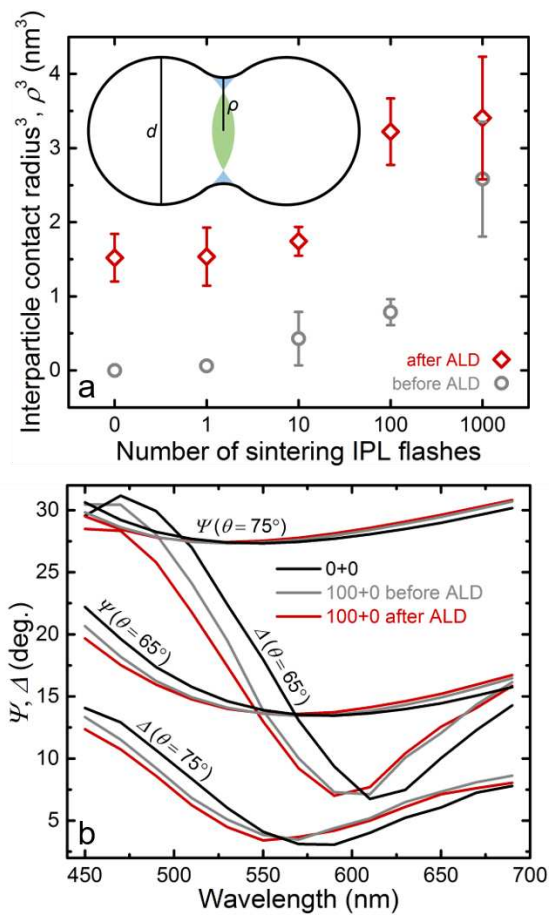

**Figure 2.** (a) $\rho$ calculated using the geometrical model described in Note 1 in the Supporting Information. Error bars represent the 95% confidence intervals based on ellipsometry precision (see Figure S3 in the Supporting Information). The "after ALD" values (red diamonds) are obtained by thermally annealing the samples after IPL to simulate the additional sintering that occurs during ALD. Since conductivity measurements are conducted after ALD, all $\rho$ values reported and used in this work are the "after ALD" values, unless specified otherwise. (b) Representative ellipsometry data used to determine ZnO volume fraction, $\phi$, which is the primary input in the geometrical model used to calculate $\rho$. $\Theta$ is the angle of incidence. $\Psi$ and $\Delta$ are measured before any treatment (black), then after sintering IPL (in this case 100 flashes, gray), and then after thermal annealing to simulate ALD (red).

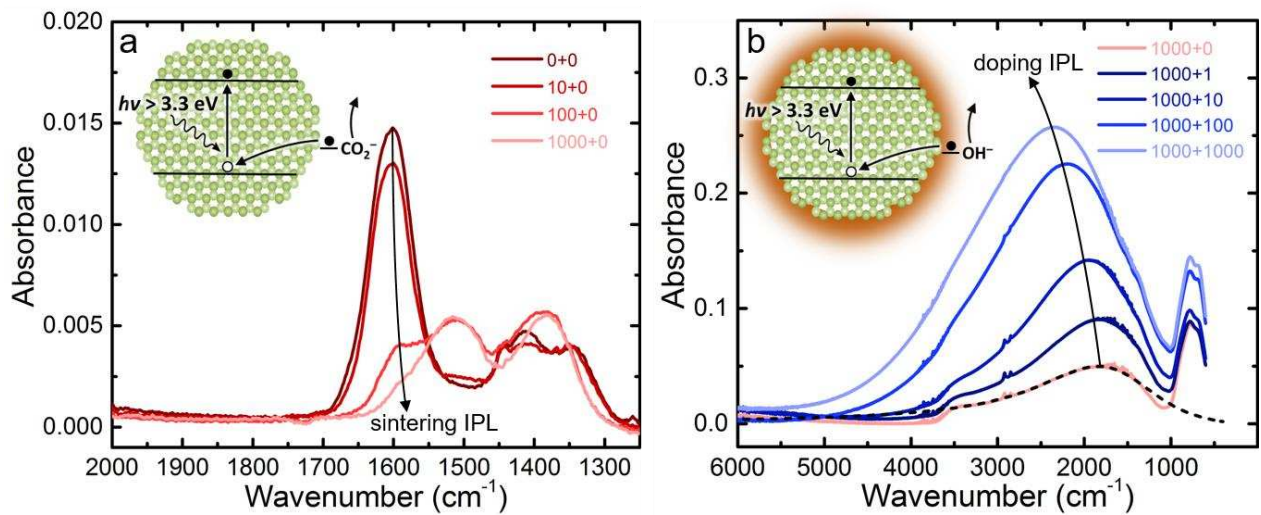

**Figure 3.** FTIR absorption spectra evolution with IPL dose. Sample names are of the form "$x+y$", where $x$ is the number of sintering IPL flashes and $y$ is the number of doping IPL flashes. (a) Sintering series: spectra after sintering IPL, collected before ALD. Surface carboxylates (1600 cm$^{-1}$) desorb after oxidization by photogenerated holes, which likely facilitates the increase in $\rho$. (b) Doping series: spectra after 1000 flashes of sintering IPL plus flashes of doping IPL, collected after ALD. Hydroxyl groups (~3500 cm$^{-1}$) are removed via oxidation by photogenerated holes, and $n$ increases from 1.1 to 7.6 × 10$^{20}$ cm$^{-3}$. The broad feature that intensifies and blueshifts from ~1800 to ~2400 cm$^{-1}$ is due to the LSPR. The dashed black line shows the fit of the "1000+0" spectrum from Equation 2.

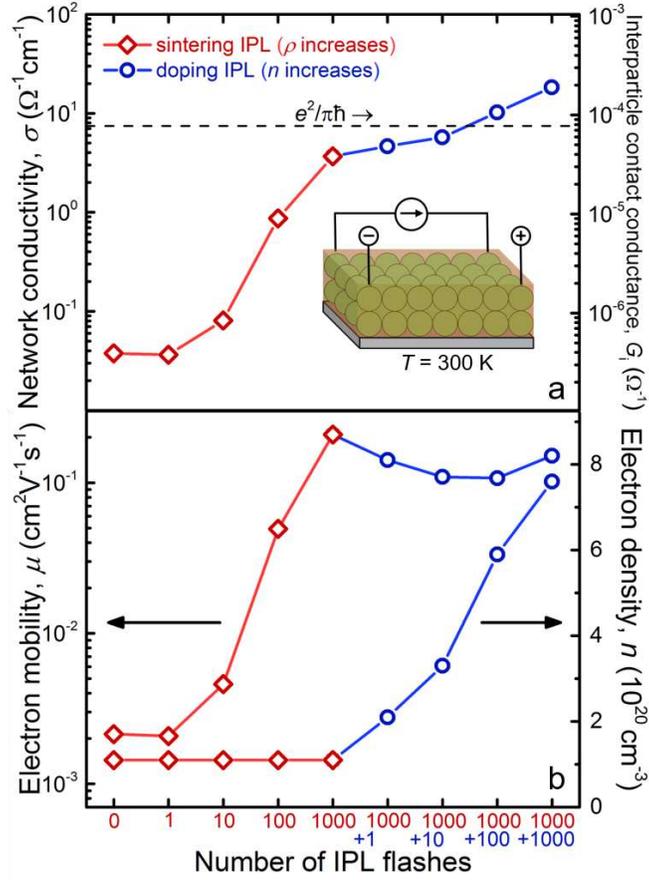

**Figure 4.** Room-temperature $\sigma$, $\mu$, and $n$ vs. number of IPL flashes. Red diamonds indicate treatment with sintering IPL only (sintering series), and blue circles indicate treatment with doping IPL after 1000 flashes of sintering IPL (doping series). (a) Left axis: $\sigma$ from the Van der Pauw method. Right axis: $G_i$, from Equation 3. (b) Right axis: $n$ from LSPR fits. Left axis: $\mu = \sigma/en$.

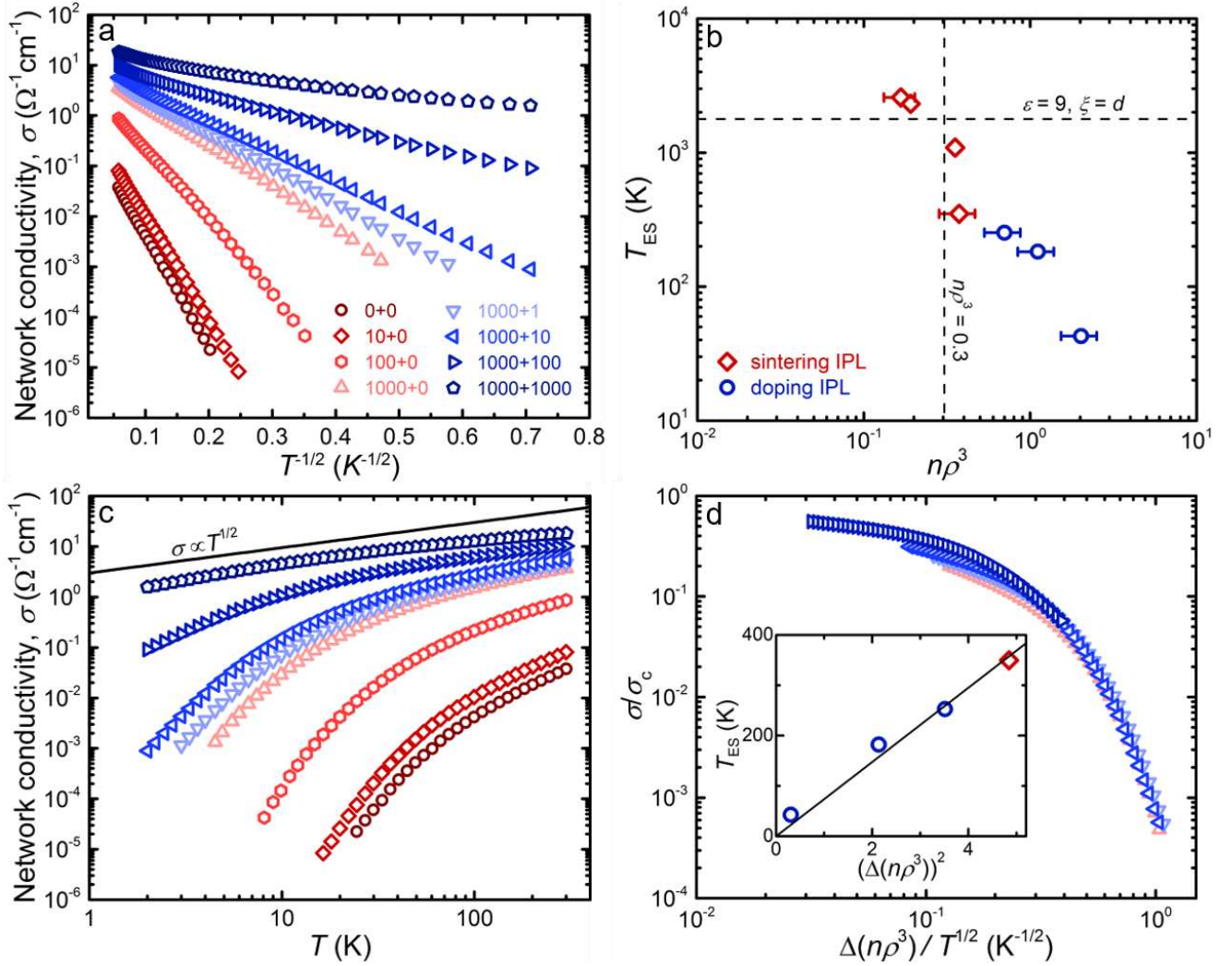

**Figure 5.** Temperature dependence of $\sigma$. Sample names are of the form "$x+y$", where $x$ is the number of sintering IPL flashes and $y$ is the number of doping IPL flashes. (a) Semilog plot of $\sigma$ vs. $T^{-1/2}$. Zabrodskii analysis of these data (Figure S8 in the Supporting Information) indicates ESVRH for all but the most conductive sample ("1000+1000"). (b) Log-log plot of $T_{ES}$ vs. $n\rho^3$. $T_{ES}$ is extracted from the slope of $\log \sigma$ vs. $T^{-1/2}$ using Equations 4 and 5. Error bars represent the 95% confidence intervals based on the precision of $\rho$ measurement (See Figure S3 in the Supporting Information) (c) Log-log plot of $\sigma$ vs. $T$ showing the approach to the power-law behavior expected at the IMT. The black line shows the slope corresponding to $\sigma \propto T^{1/2}$. (d) Log-log plot of $\sigma/\sigma_c$ vs. $\Delta(n\rho^3)/T^{1/2}$ of the four most conductive samples showing the scaling behavior expected near the IMT (see Equation 6). Inset: corresponding $T_{ES}$ vs. $[\Delta(n\rho^3)]^2$.

ASSOCIATED CONTENT

**Supporting Information.** The Supporting Information is available free of charge on the ACS Publications website at DOI: XX.XXXX/acs.nano-lett.XXXXXXX. Results of thermal annealing experiments; XRD patterns and SEM images; fits of ellipsometry data and $\phi$ and $t$ values; results of low-intensity illumination experiments; evolution of carboxylate features in FTIR spectra; Hall effect data; table of $\rho$, $n$, and $\sigma$ of all samples; LSPR of sample not treated with IPL; Zabrodskii plot; geometrical model used to compute $\rho$ from $\phi$; discussion of hydroxyl removal during doping IPL; theoretical estimate of $\xi$ from $n\rho^3$; (PDF)

AUTHOR INFORMATION

**Corresponding Authors.** *Email: kortshagen@umn.edu, gree1729@umn.edu

**Author Contributions.** B.L.G. conceived and executed most of the experiments. Z.L.R. and B.N.V. performed electron transport measurements and analyzed the results. C.G. performed SEM and XRD measurements and analyzed the results. K.V.R., L.F.F., and B.I.S. contributed to data interpretation. E.S.A. and U.R.K. supervised the project and contributed to data interpretation. All authors participated in preparation of the manuscript.

**Notes.** The authors declare no competing financial interest.

ACKNOWLEDGEMENTS

This work was supported primarily by the National Science Foundation through the University of Minnesota MRSEC under Award Number DMR-1420013. Parts of this work were carried out in the College of Science and Engineering Characterization Facility, University of Minnesota, which has received capital equipment funding from the NSF through the UMN MRSEC program under Award Number DMR-1420013. Parts of this work were carried out in the College of Science and Engineering Minnesota Nanocenter, University of Minnesota, which receives partial support from the NSF through the NNIN program. We thank Tim Peterson for assistance with electron transport measurements, Bryce Williams for assistance with intense pulsed light annealing, Tom Fielitz for assistance with ellipsometry, and Prof. Alexander Govorov for helpful discussions.

TOC GRAPHIC

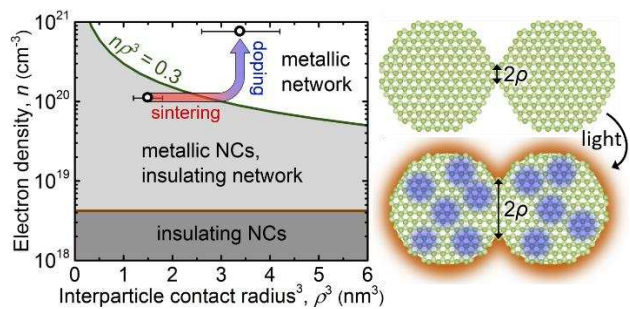

Supporting Information for

# ZnO Nanocrystal Networks Near the Insulator-Metal Transition: Tuning Contact Radius and Electron Density with Intense Pulsed Light

Benjamin L. Greenberg,[a] Zachary L. Robinson,[b] K. V. Reich,[b,d] Claudia Gorynski,[e]
Bryan N. Voigt,[c] Lorraine F. Francis,[c] B. I. Shklovskii,[b] Eray S. Aydil,[c] and Uwe R. Kortshagen[a]

[a]Department of Mechanical Engineering, [b]Department of Physics,
and [c]Department of Chemical Engineering and Materials Science,
University of Minnesota, Minneapolis, Minnesota 55455, United States

[d]Ioffe Institute, St Petersburg, 194021, Russia

[e]Department of Mechanical and Process Engineering,
University of Duisburg-Essen, Forsthausweg 2, 47057 Duisburg, Germany

Email: kortshagen@umn.edu, bengreenberg@umn.edu

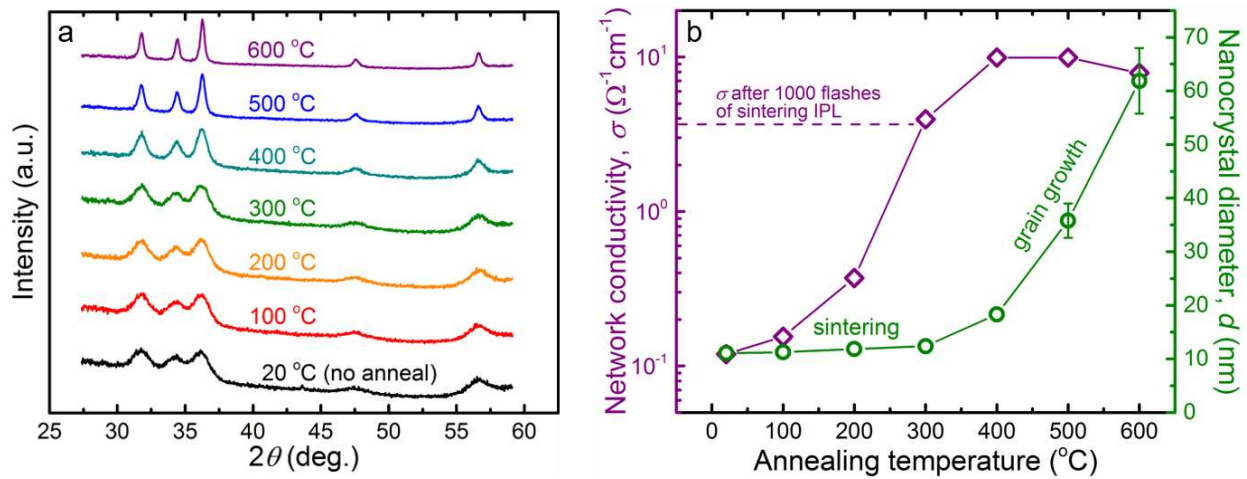

**Figure S1.** ZnO NC networks on glass substrates (for conductivity measurements) and Si substrates (for XRD) were thermally annealed to determine the temperature range in which rapid grain growth occurs. The samples were heated in air for 11 minutes (the duration of a 1000-flash IPL treatment). (a) XRD patterns show the onset of significant grain growth at 400 °C. (b) Right axis: $d$ from the Scherrer formula for spheres.[1] Left axis: room-temperature $\sigma$ from the Van der Pauw method. At 300 °C, $\sigma$ increases by a factor of ~30 (and is approximately equal to that of a sample treated with 1000 flashes of sintering IPL), while $d$ remains nearly constant. This is consistent with entering the initial stage of sintering. As discussed in the text and shown in Figure S3, IPL increases film density only slightly (by ~6%), which is consistent with remaining in the initial stage rather than proceeding to the intermediate densification stage.[2]

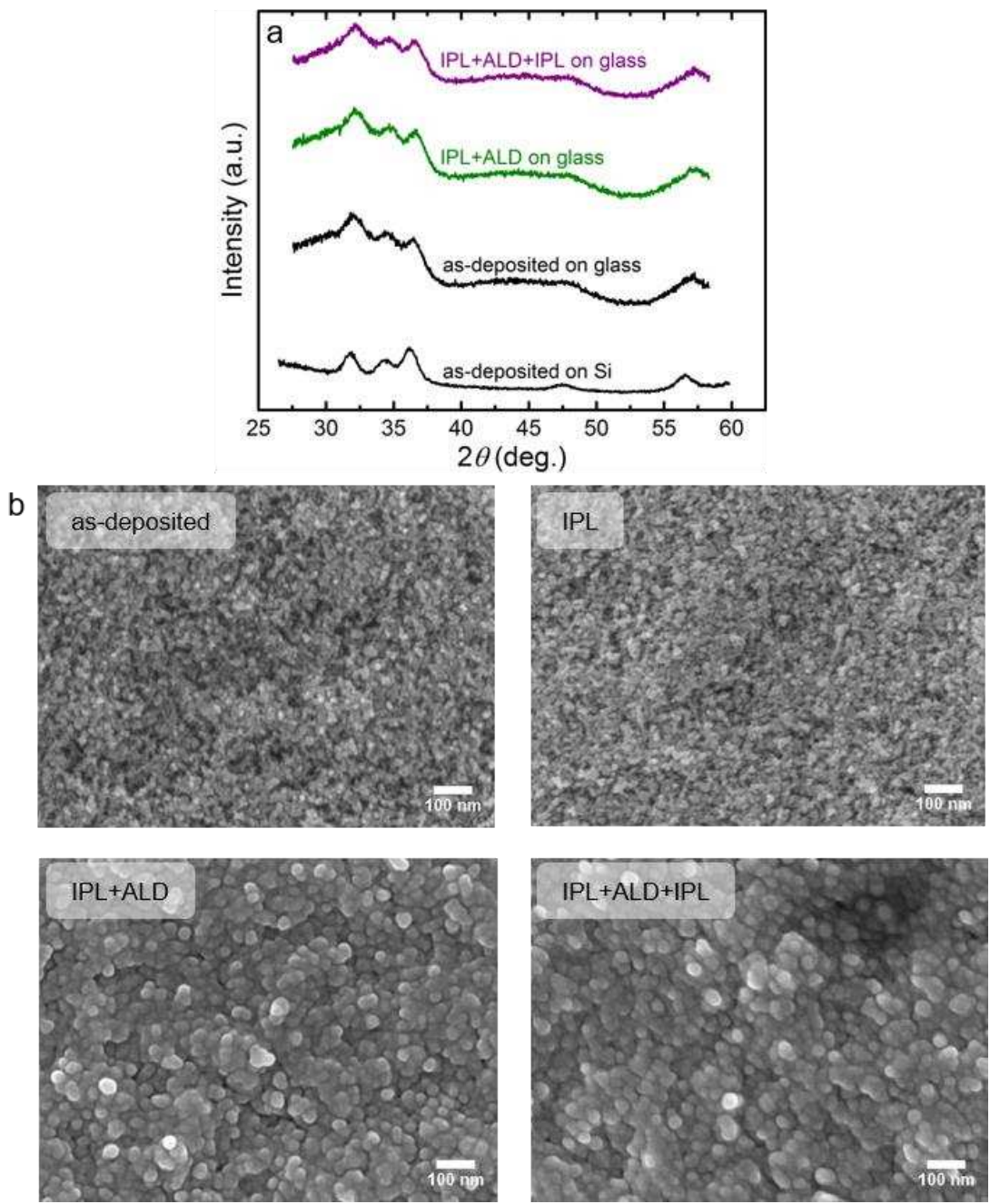

**Figure S2.** (a) XRD patterns of ZnO NCs at various treatment stages. NCs were deposited (black), then sintered by IPL and coated by ALD (green), then photodoped by IPL (purple). These treatments do not significantly change the NC diameter. Here the number of flashes was 1250 at both stages of IPL. Since thermal behavior during IPL is substrate dependent, the substrates used were the same as those of samples that undergo electrical measurements (boro-aluminosilicate glass). To reduce the background signal and facilitate peak fitting, an additional XRD pattern was acquired from as-deposited ZnO NCs on a Si substrate. This is the pattern used to determine via the Scherrer formula for spheres[1] that the average NC diameter is $9.6 \pm 0.9$ nm. (b) Corresponding plan-view SEM micrographs. Neither IPL before ALD nor IPL after ALD cause significant structural changes visible at this magnification. Scale bars represent 100 nm.

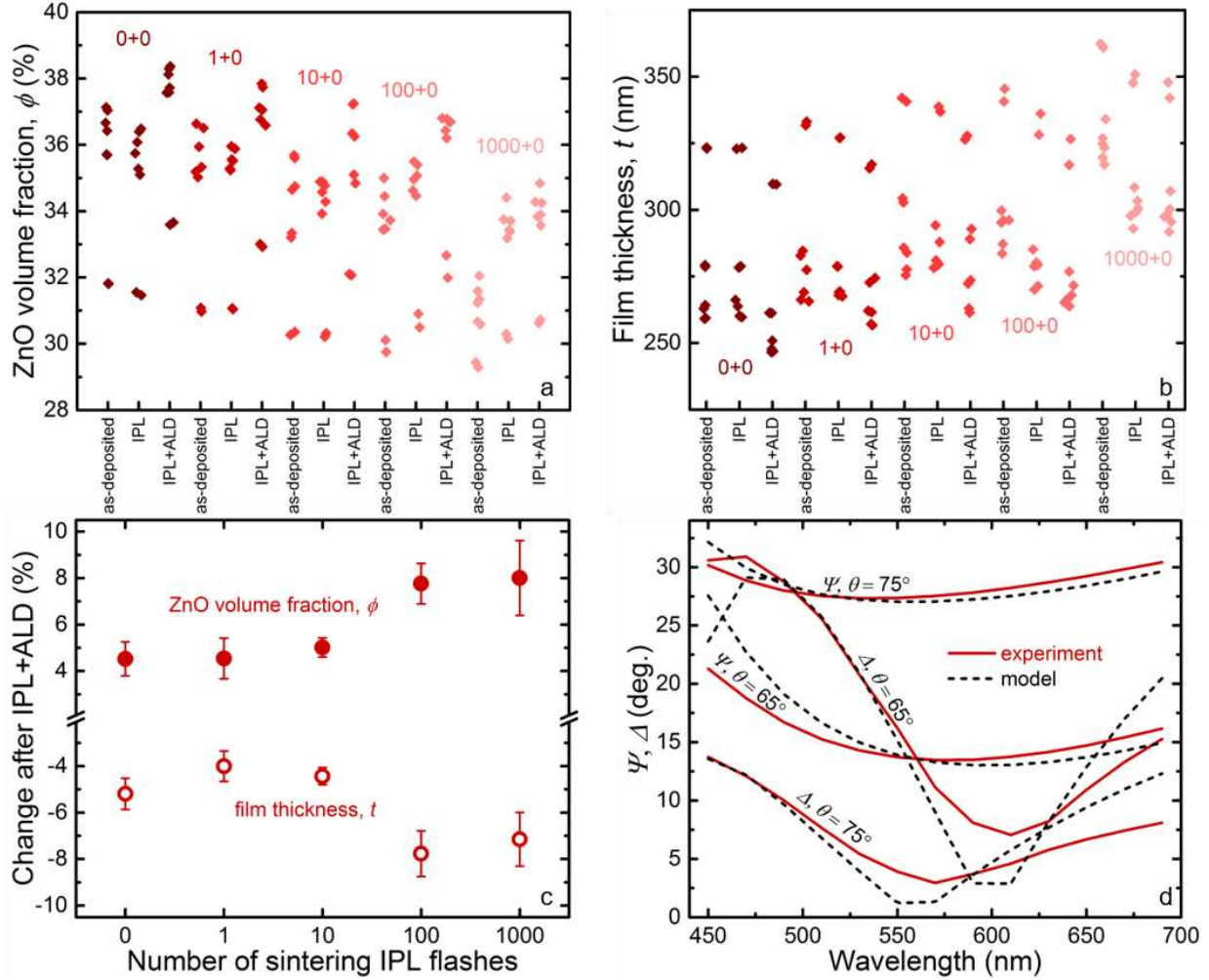

**Figure S3.** (a,b) ZnO volume fraction, $\phi$, and film thickness, $t$, from ellipsometry at various treatment stages. The $\phi$ values are used to determine the $\rho$ values shown in Figure 2 in the text. Five ZnO NC films were deposited simultaneously (for five different values of the number of IPL flashes) and each film was measured in eight locations at each treatment stage, so that each data column contains eight points (some points overlap). "ALD" refers to the thermal treatment described in the text which simulates the thermal conditions of ALD without $Al_2O_3$ deposition. The increase in $\rho$ due to sintering during IPL and ALD are computed using pairs of $\phi$ values measured at the same location of the same sample. Due to statistical fluctuations, some pairs showed a slight decline in $\phi$ after 10 or fewer IPL flashes; these data are not used in calculating $\rho$, because the geometrical model does not apply in these cases. See Note 1 for details. We note that the trends in the absolute values of $\phi$ and $t$ may initially appear inconsistent with sintering ($\phi$ is lower and $t$ is higher in samples treated with more flashes). In actuality, these trends, as well as scatter within each sample, are due to spatial nonuniformity of the NC deposition which produces variance in *as-deposited* $\phi$ corresponding to variance in NC packing rather than in $\rho$. To confirm that this variance in packing does not significantly impact our results, we also applied 1000 IPL flashes and "ALD" to a sample with spatially averaged as-deposited $\phi \approx 34\%$, and found that after treatment, $\rho = 1.5 \pm 0.1$ nm, equal to that of the sample which received the same treatment but had as-deposited $\phi \approx 31\%$. (c) Spatially averaged $\phi$ and $t$ measured after IPL and ALD and normalized to their as-deposited values. Here sintering due to IPL and ALD is more immediately apparent. (d) Representative ellipsometry data and two-parameter fit of $\Psi$ and $\Delta$ used to determine $\phi$ and $t$. Fits are produced using the Bruggeman effective medium approximation.

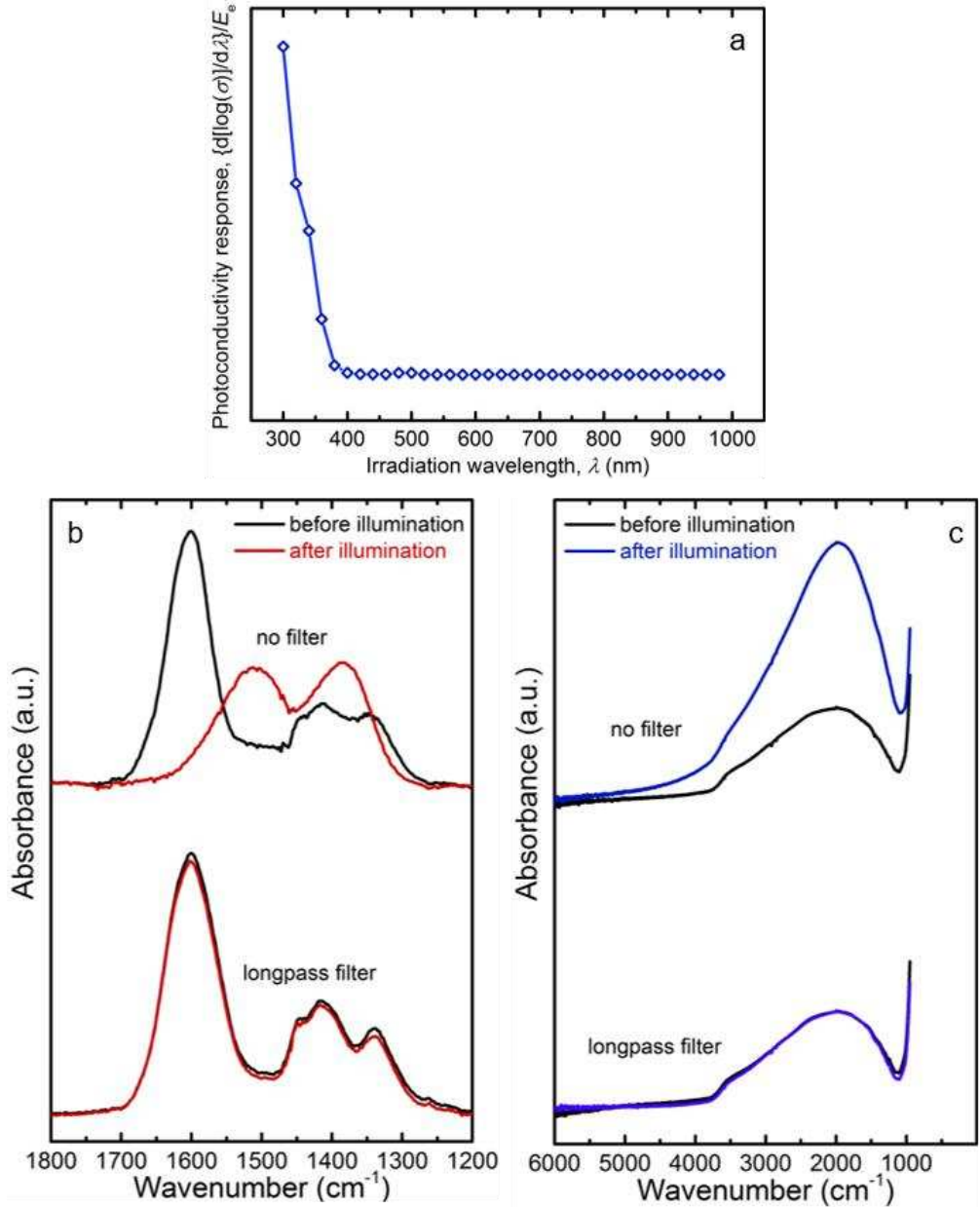

**Figure S4.** Results of low-intensity illumination experiments. ZnO NC networks were irradiated with a xenon arc lamp under ambient conditions at a spectrally integrated surface power density of approximately 200 mW/cm² so that they remain close to room temperature. (a) Photoconductivity spectrum of a ZnO NC network acquired after ALD treatment using monochromated lamp emission. (b,c) FTIR spectra before and after illumination, with and without a longpass filter with a cut-on wavelength of 450 nm. The total photon dose is equivalent to 100 flashes of IPL. (b) Carboxylate and related features of an as-deposited sample (no ALD or IPL). Unfiltered light has the same effect as sintering IPL (see Figure 3a in the text). Longpass-filtered light has no significant effect. (c) Hydroxyl and LSPR features of samples pre-treated with 1000 flashes of sintering IPL followed by ALD to mimic the doping IPL series. Qualitatively, unfiltered light has the same effects as doping IPL, revealing that photochemical hydroxyl removal and $n$ enhancement can occur near room temperature, although quantitatively these effects are diminished (see Note 2 for further discussion).

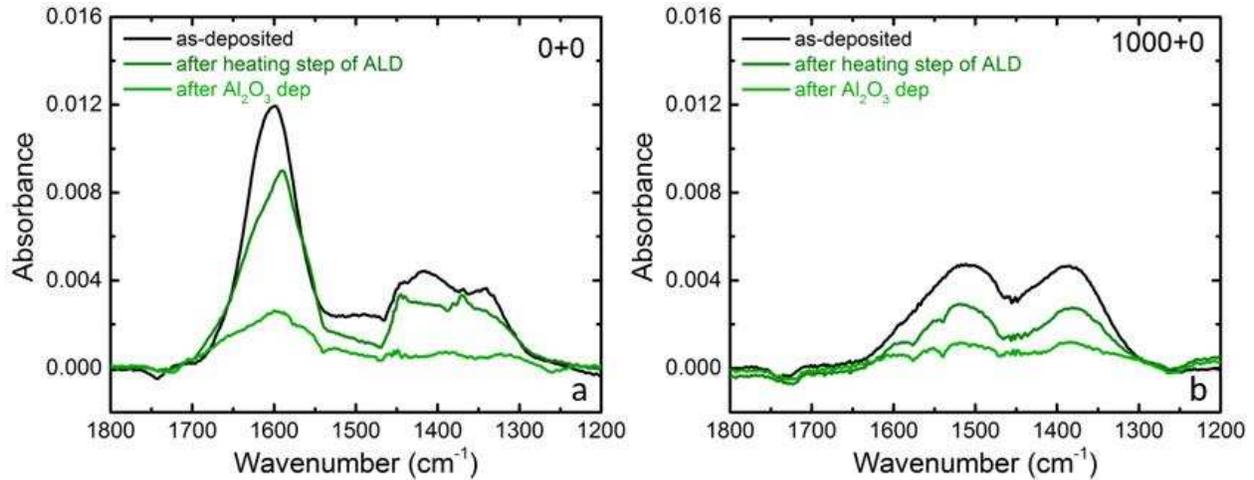

**Figure S5.** Evolution of carboxylate features in FTIR absorption spectra due to ALD. Spectra were acquired before ALD (black), then after the initial heating step of ALD which entails heating at 180 °C under $N_2$ for 33 minutes (dark green), and then after ALD of 7.7 nm of $Al_2O_3$ was completed (light green). (a) In the case of no IPL treatment ("0+0"), the as-deposited spectrum shows carboxylate features at 1600 cm⁻¹ (asymmetric stretch, $v_{as}(COO^-)$) and between 1300 and 1500 cm⁻¹ (symmetric stretch, $v_s(COO^-)$, among other modes). After the heating step of ALD (during which sintering occurs), the $v_{as}(COO^-)$ stretch is shifted to 1591 cm⁻¹, indicating a change in the form of carboxylate binding to the ZnO NC surfaces. A possible explanation of the redshift is a change from ionic monodentate bonding to a covalent bridging configuration, where a carboxylate group is bound to two Zn atoms.[3] Such a configuration could be formed by a carboxylate group binding to two adjacent ZnO NCs during sintering and thus interfering with $\rho$ enhancement. We cannot rule out other transformations (such as ionic monodentate to covalent monodentate), but in any case, the persistence and evolution of the carboxylate features indicate that carboxylate groups play a role in surface chemistry changes during sintering. The carboxylate features also remain after the deposition step of ALD. The blueshift of the $v_{as}(COO^-)$ stretch during this stage could be due to another binding reconfiguration and/or interaction between carboxylates and $Al_2O_3$. (b) In the case of 1000 IPL flashes before ALD ("1000+0"), the carboxylate features are completely removed during IPL. We assign the features at 1512 and 1388 cm⁻¹ to carbonate/water complexes, which form from reactions between $CO_2$ and OH groups on metal oxide surfaces.[4–6]

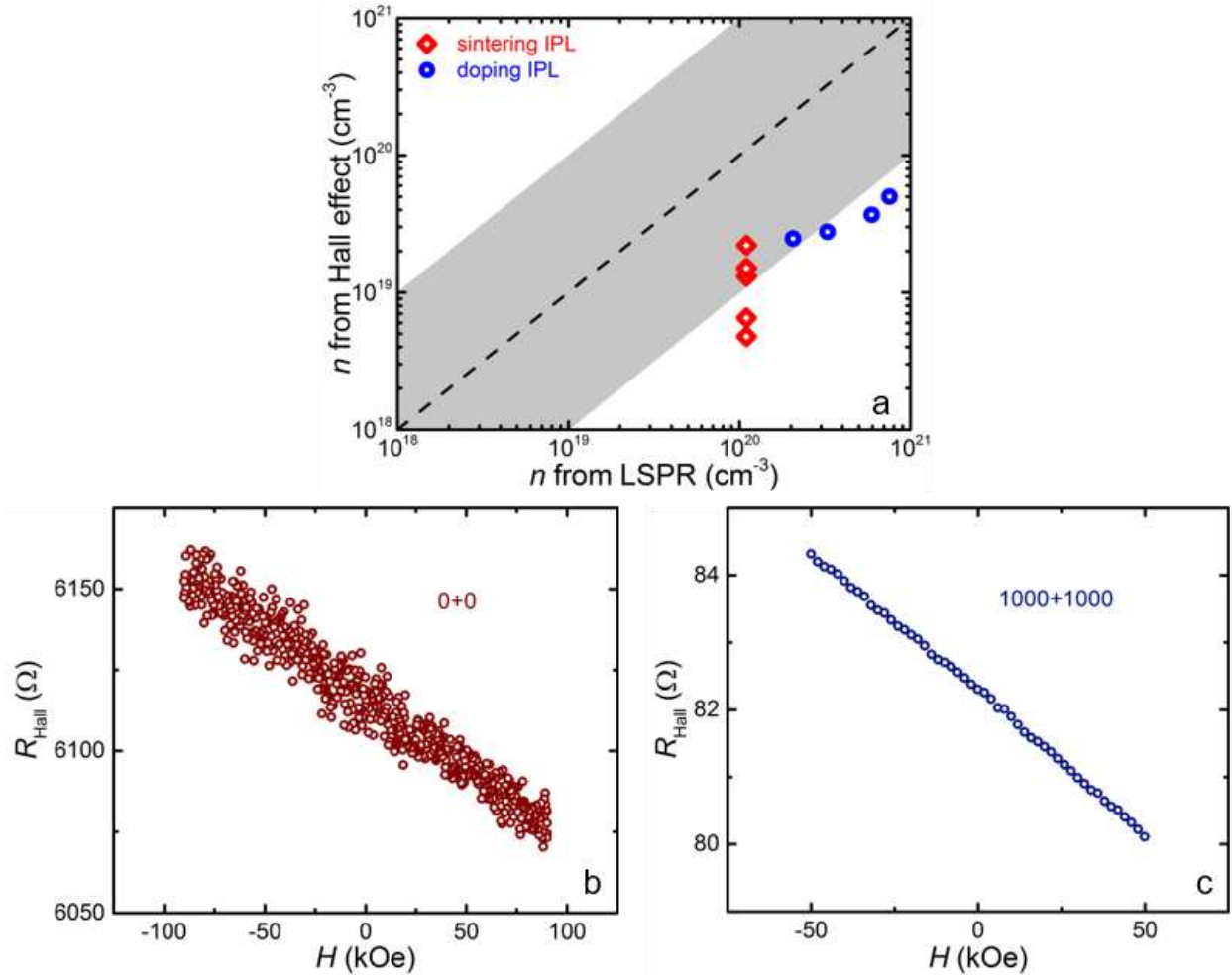

**Figure S6.** (a) $n$ from Hall effect measurements vs. $n$ from the LSPRs shown in figure 3b in the text. The gray region represents agreement within an order of magnitude. (b,c) Hall data from the least and most conductive samples, "0+0" and "1000+1000". For resistive samples such as "0+0", $R_{Hall}$ drifted (decreased) significantly over time at zero field in the low-pressure measurement chamber. The $R_{Hall}$ drift rate, $dR/dt$, was constant, so we deconvolved $dR/dt$ and $dR/dH$ by sweeping the magnetic field in both directions. Each $R_{Hall}$ value plotted in (b) is an average of the two corresponding values obtained from the two sweeps.

**Table S1.** $\rho^3$, $n$, and $\sigma$ of all samples. $\rho^3$ is determined by ellipsometry (Figure 2 in the text), $n$ is determined from the LSPR absorption features in the FTIR spectra (Figure 3 in the text), and $\sigma$ is determined by four-point probe measurements using the Van der Pauw geometry (Figure 4 in the text). Also tabulated are $\sigma$ of a sample treated with 2000 IPL flashes before ALD and a sample treated with 2000 IPL flashes after ALD. Comparing $\sigma$ of these samples to $\sigma$ of the "1000+1000" sample (which also received a total of 2000 flashes) confirms that IPL before ALD and IPL after ALD increase $\sigma$ by different mechanisms (shown in the text to be sintering and doping), and that a combination of the two treatments is necessary for maximizing $\sigma$.

| sample name | IPL flashes before ALD (sintering IPL) | IPL flashes after ALD (doping IPL) | interparticle contact radius[3], $\rho^3$ (nm$^3$) | electron density, $n$ ($10^{20}$ cm$^{-3}$) | network conductivity, $\sigma$ ($\Omega^{-1}$cm$^{-1}$) |
|---|---|---|---|---|---|
| 0+0 | 0 | 0 | 1.5 ± 0.3 | 1.1 | $3.8 \times 10^{-2}$ |
| 1+0 | 1 | 0 | 1.5 ± 0.4 | 1.1 | $3.7 \times 10^{-2}$ |
| 10+0 | 10 | 0 | 1.7 ± 0.2 | 1.1 | $8.1 \times 10^{-2}$ |
| 100+0 | 100 | 0 | 3.2 ± 0.4 | 1.1 | $8.7 \times 10^{-1}$ |
| 1000+0 | 1000 | 0 | 3.4 ± 0.8 | 1.1 | $3.7 \times 10^{0}$ |
| 1000+1 | 1000 | 1 | 3.4 ± 0.8 | 2.1 | $4.7 \times 10^{0}$ |
| 1000+10 | 1000 | 10 | 3.4 ± 0.8 | 3.3 | $5.7 \times 10^{0}$ |
| 1000+100 | 1000 | 100 | 3.4 ± 0.8 | 5.9 | $1.0 \times 10^{1}$ |
| 1000+1000 | 1000 | 1000 | 3.4 ± 0.8 | 7.6 | $1.8 \times 10^{1}$ |
| -- | 2000 | 0 | -- | -- | $5.0 \times 10^{0}$ |
| -- | 0 | 2000 | -- | -- | $3.6 \times 10^{0}$ |

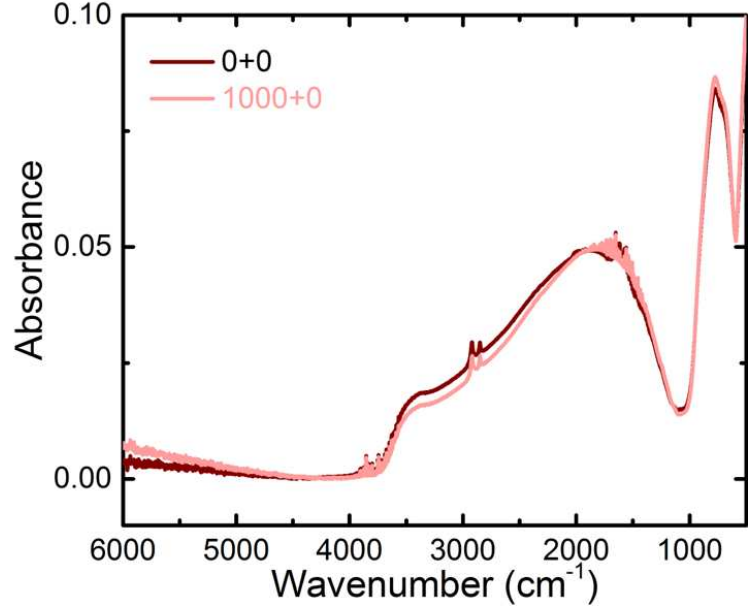

**Figure S7.** FTIR spectra, acquired after ALD, of a sample treated with 1000 flashes of sintering IPL ("1000+0") and a sample not treated with IPL ("0+0"). The sintering IPL has no significant effect on the LSPR, indicating that $n$ remains constant.

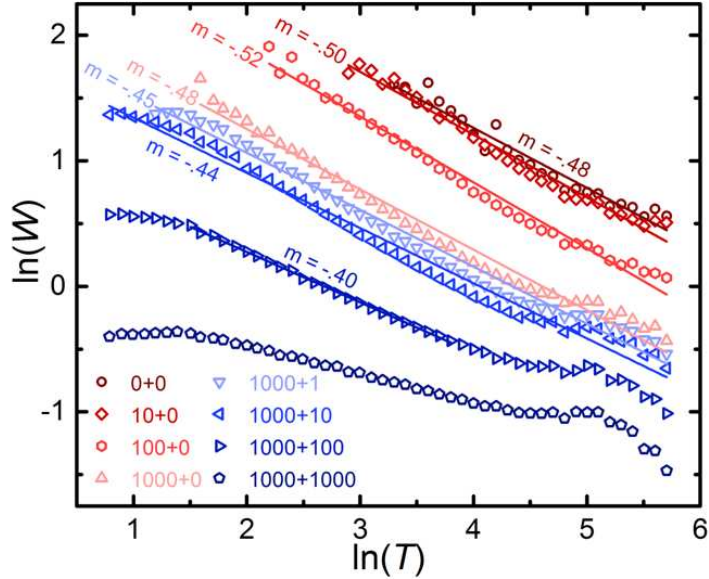

**Figure S8.** Zabrodskii plot of ln $W$ vs. ln $T$, where $W = \mathrm{d}(\ln(\sigma))/\mathrm{d}(\ln(T))$. Slopes near 1/2 are consistent with ESVRH. The flattening at low $T$ shown by the two most conductive samples ("1000+100" and "1000+1000") is consistent with a transition to a power law.

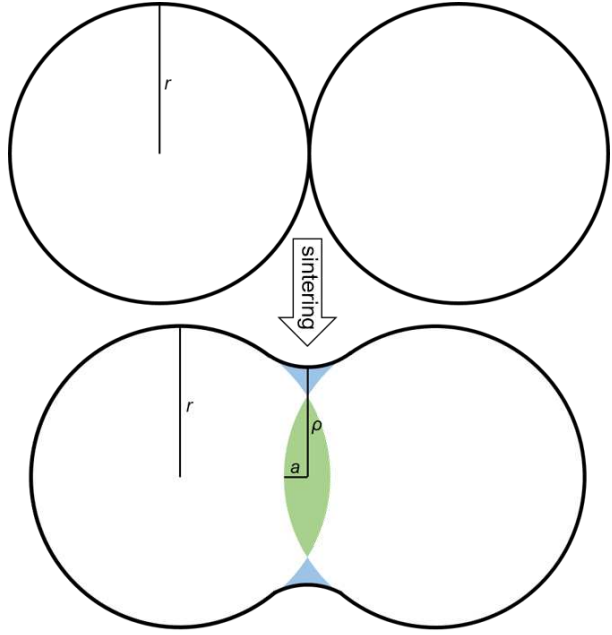

**Figure S9.** Geometrical model used to compute $\rho$ from $\phi$.

**Note 1.** Geometrical model used to compute $\rho$ from $\phi$.

ZnO volume fractions from ellipsometry, $\phi$, are converted to estimates of interparticle contact radii, $\rho$, using a simple geometrical model. We assume $\rho_0 = 0$, and then we calculate changes in $\rho$ from changes in $\phi$. Our model begins with a network of spheres of radius $r$ and packing fraction $\phi_0$ that are connected by point contacts. Consider a pair of connected spheres within this network. The initial center-to-center distance is $2r$. During sintering, the packing fraction increases to $\phi$ and the center-to-center distance decreases to $2(r - a)$, where $a$ and $\phi$ are related by

$$\frac{\phi}{\phi_0} \approx \left(\frac{r}{r-a}\right)^3 \tag{1}$$

This reduction of center-to-center distance displaces a volume of material (green region), given by

$$V_{\text{disp}} = \frac{2\pi}{3}a^2(3r - a) \tag{2}$$

Then curvature is minimized when this displaced material takes the form of the solid of revolution obtained by rotating the region bounded by the NC cross-sections and a circle tangent to the NCs (blue region) about the axis connecting the NC centers. The minimum outer radius of this solid is $\rho$, which we can find be equating the volume of the solid of revolution, $V_{\text{sol}}$, with $V_{\text{disp}}$, i.e. by equating the volume of the blue region with the volume of the green region. To obtain an expression for $V_{\text{sol}}$ we use a Cartesian coordinate system with origin at the midpoint between the NC centers, so that the NC cross-sections have centers at $(\pm(r - a), 0)$, and the circle forming the upper bound of the blue region and tangent to the NCs at the horizontal coordinates $\pm x_{\text{int}}$ has radius $r_{\text{UB}}$ and center at $(0, k_{\text{UB}})$, where

$$r_{\text{UB}} = \sqrt{x_{\text{int}}^2 + \left(k_{\text{UB}} - \sqrt{r^2 - (x_{\text{int}} + r - a)^2}\right)^2} \qquad (3)$$

$$k_{\text{UB}} = \sqrt{r^2 - (x_{\text{int}} + r - a)^2} \left(1 - \frac{x_{\text{int}}}{x_{\text{int}} + r - a}\right) \qquad (4)$$

Now we equate $V_{\text{sol}}$ with $V_{\text{disp}}$ and solve for $x_{\text{int}}$:

$$\pi \int_{-x_{\text{int}}}^{x_{\text{int}}} \left[\left(k_{\text{UB}} - \sqrt{r_{\text{UB}}^2 - x^2}\right)^2 + (x + r - a)^2 - r^2\right] \mathrm{d}x = V_{\text{disp}} \qquad (5)$$

Finally, $\rho$ is given by

$$\rho = k_{\text{UB}} - r_{\text{UB}} \qquad (6)$$

**Note 2.** Fate of hydroxyls during doping IPL.

FTIR spectra acquired after doping IPL (Figure 3b) clearly show elimination of the hydroxyl signal and enhancement of the LSPR signal, but they show no other significant changes. This leaves open the question of the ultimate fate of oxygen and hydrogen atoms after hydroxyl removal. Interactions among hydroxyls, UV photons, and photogenerated electrons/holes may produce a wide variety of chemical species, including neutral OH, atomic O and H, diatomic $O_2$ and $H_2$, and $H_2O$. Broadly speaking, the FTIR spectra allow two possibilities: (1) the species formed are in fact exclusively undetectable via FTIR (IR inactive or active outside the wavenumber range of the measurement) or (2) some or all of the species diffuse out of the film. The latter is perhaps counterintuitive, given that the ZnO NC networks treated with doping IPL are first infilled with $Al_2O_3$, and that we have proposed that hydroxyl groups are initially located at or near the ZnO/$Al_2O_3$ interfaces (so that they interact with charge carriers originating from the ZnO). Still, appreciable diffusion of chemical species out of the films under UV irradiation is a possibility worth considering in more detail.

To understand diffusion through the $Al_2O_3$-coated ZnO NC networks, it is useful to consider the ALD process by which they are formed. The ALD infilling is a self-limiting process that decreases the porosity of the film until it reaches a value at which the ALD precursors can no longer infiltrate the film under the pulse conditions (temperature = 180 °C, precursor partial pressure ≈ 100 Torr, time = 0.1 s). This minimum porosity was shown to be ~5-10% for ZnO ALD on ZnO NCs by Thimsen *et al.*,[7] and the value for $Al_2O_3$ ALD is likely similar. Because these pores are what enable ALD precursor infiltration, they likely form continuous or nearly continuous networks from the top to the bottom of the film, i.e. pore networks with a permeability just slightly below that which would enable further deposition under the pulse conditions. These pore networks can therefore later serve as diffusion pathways for species entering and exiting the film. (Indeed, the air stability of these samples is measurably imperfect: whereas conductivity shows no decay after months of storage in the dark under an inert atmosphere, it decays at a rate of ~1% per day under dark ambient conditions, presumably due to room-temperature diffusion of $H_2O$ and/or $O_2$ into the film.) During IPL, when a molecule is generated at a ZnO/$Al_2O_3$ interface, the rate-determining step for diffusion out of the film is likely passage through a fairly short length of $Al_2O_3$ in order to reach a pore network. On average this length is probably no longer than ~7-8 nm, which is the thickness of a 70-cycle $Al_2O_3$ ALD coating on a flat substrate.

The question then becomes whether, under the irradiation conditions, the diffusion length of a photogenerated species through the $Al_2O_3$, $l$, is 7-8 nm. Of particular interest is the low-intensity illumination experiment (Figure S4c), in which the film remains close to room temperature. In this experiment, the elapsed time is ~6000 s, and for the sake of argument we can assume that the diffusing species is $H_2O$. In order to obtain $l$ = 7-8 nm, we require $D_{300K} \approx 10^{-17}$ cm²/s for $H_2O$. Generally, diffusion coefficients in $Al_2O_3$ are measured at temperatures well above 300 K, but still we can use the results of these experiments to gauge plausibility. For $H_2O$ diffusing in polycrystalline Lucalox™ $Al_2O_3$, ~$10^{-17}$ cm²/s is the value of $D_{600K}$.[8] Therefore we require that the activation energy for $H_2O$ diffusion in our $Al_2O_3$ is approximately half that in Lucalox (assuming comparable $D_0$). This is in fact not unreasonable, considering that $Al_2O_3$ grown on ZnO NCs by ALD at 180 °C is amorphous and likely has relatively high densities of coordination defects and impurities due to substrate effects, low growth temperature, and possible chamber contamination. In general, diffusion activation energies vary widely among various forms of $Al_2O_3$; for example, activation energies for oxygen diffusion in amorphous and $\gamma$-$Al_2O_3$ have been shown to be five times lower than that in $\alpha$-$Al_2O_3$.[9]

To conclude, we cannot rule out either of the potential explanations of the FTIR spectra: the chemical species formed after photooxidation of hydroxyls may be difficult to detect, capable of diffusing out of the film, or both. Any of these possibilities is plausible and compatible with main conclusions presented in the text. Future studies aimed at distinguishing among them should include further analysis of impurities in the films as well as any gases evolved during irradiation.

**Note 3**. Estimate of $\xi$ from $n\rho^3$ vs. $\xi$ from $T_{ES}$.

It has been shown[10] that $\xi$ in a network of NCs with diameter $d$, free electron density $n$, and interparticle contact radius $\rho$ can be estimated from

$$\xi = A \frac{d}{\ln(2/n\rho^3)}, \tag{7}$$

provided that (1) the network is far from the IMT on the insulating side and at low temperature so that electron hopping proceeds via elastic cotunneling through chains of NCs and (2) the number of free electrons per NC is greater than $\alpha^{-3/2}$, where $\alpha = 2\delta d/d$. $A$ is a constant of order unity. Applying Equation 7 to our sample farthest from the IMT ("0+0", $n\rho^3 = 0.17$) and assuming $A = 1$, we obtain $\xi = 4$ nm. For comparison, using Equation 5 in the text and $T_{ES} = 2600$ K while assuming a dielectric constant of 9, we obtain $\xi = 7$ nm. Equation 7, however, should not be applied to samples approaching the IMT, because it is invalid when $\xi > d$ (for details, see ref. 10). Hence we use scaling theory to model the dependence of $T_{ES}$ on $n\rho^3$ near the IMT, as shown in Figure 5d in the text.